\documentclass[aps,prd,reprint,nofootinbib,superscriptaddress,longbibliography]{revtex4-2}

\usepackage{amsmath,amssymb,bm,mathtools}
\usepackage{graphicx}
\usepackage{booktabs}
\usepackage{array}
\usepackage{microtype}
\usepackage{xcolor}
\usepackage[colorlinks=true,citecolor=blue!55!black,linkcolor=blue!55!black,urlcolor=blue!55!black]{hyperref}

\newcommand{\dd}{\mathrm{d}}
\newcommand{\Lie}{\mathcal{L}}
\newcommand{\eps}{\boldsymbol{\epsilon}}
\newcommand{\ThetaLW}{\boldsymbol{\Theta}}
\newcommand{\Q}{\boldsymbol{Q}}
\newcommand{\kform}{\boldsymbol{k}}
\newcommand{\omegaf}{\boldsymbol{\omega}}
\newcommand{\cE}{\mathbb{E}}

\begin{document}

\title{Hair or a Boundary Source? Covariant Phase Space of a Static Type-I Black Hole}
\author{Yi-kun Li}
\email{liyikun@xao.ac.cn}
\affiliation{State Key Laboratory of Radio Astronomy and Technology,
Xinjiang Astronomical Observatory, CAS, 150 Science 1-Street,
Urumqi 830011, China}
\affiliation{School of Astronomy and Space Science,
University of Chinese Academy of Sciences, No.19A Yuquan Road,
Beijing 100049, China}

\begin{abstract}
We study the two-parameter static type-I vacuum black hole recently obtained
from an electromagnetic seed by a nonlinear solution-generating map.  Its
parameter \(B\) changes dimensionless horizon geometry, yet the Komar mass
variation at fixed asymptotic time is incompatible with the entropy variation.
Using the Iyer--Wald, Barnich--Brandt, and Lee--Wald constructions, we derive the
surface-charge one-form throughout the regular solution space and show that it
has a nonzero curl.  The same obstruction is measured by symplectic flux
through the limiting Weyl boundary.  A field-dependent normalization of the
stationary Killing field supplies an integrating factor; the resulting energy
\(H=m/(1+B^2m^2)^{3/2}\) is uniquely selected by Schwarzschild mass
normalization within this class.  Homothetic scaling leaves the fixed-\(G\)
phase space two-dimensional because its tangent carries nonzero exact-symmetry
charges.  In Weyl coordinates, spatial infinity at finite mass is a closed
surface of revolution.  Its Brown--York canonical response reproduces the
Lee--Wald flux pointwise.  The semi-infinite annular source of the massless
background emerges as a nonuniform limit of this surface.  Finally, with
\(y=Bm\), the mechanics takes the global form
\(\dd M=T\dd S+\Psi_y\dd y\).  Thus \(B\) labels a genuine geometric
deformation while its variation acts as external-source work in the
Hamiltonian associated with fixed boundary time.
\end{abstract}

\maketitle

\section{Introduction}
\label{sec:introduction}

Stationary vacuum black holes are exceptionally rigid when asymptotic
flatness, regularity, and connected nondegenerate horizons are imposed
\cite{Israel1967,Robinson1975,ChruscielCostaHeusler2012}.  External fields and
nonstandard asymptotic regions relax the global hypotheses behind this
rigidity.  They thereby admit exact geometries whose additional parameters can
encode distortion, normalization, or boundary data distinct from an ordinary
Gauss-law charge.  Magnetized black holes provide a familiar example:
the magnetic field changes both the asymptotic frame and black-hole
thermodynamics, and the first law acquires a work term conjugate to the
external field \cite{Ernst1976,ErnstWild1976,GibbonsPangPope2014,
AstorinoCompereOliveriVandevoorde2016}.

Astorino recently found a static, Ricci-flat, algebraically general black hole
by applying an enhanced Ehlers transformation to a Schwarzschild seed
\cite{Astorino2026}.  The solution has two continuous parameters, \(m\) and
\(B\), and reduces continuously to Schwarzschild as \(B\) vanishes.  The
second parameter distorts the horizon, alters the angular period required for
regularity, and changes the Komar mass.  Its origin in an electromagnetic
seed, followed by a map to vacuum gravity, raises a basic question: does \(B\)
represent black-hole hair, or does it record the gravitational imprint of an
external source?

This question is naturally Hamiltonian.  In a covariant phase space, a
stationary Killing field generates an energy only when its surface-charge
variation is integrable over the solution manifold
\cite{LeeWald1990,IyerWald1994,BarnichBrandt2002,BarnichCompere2008}.
Unusual asymptotics often make this requirement as informative as the value of
a candidate mass itself.  Recent work on the rotating
Kerr--Bertotti--Robinson family found precisely such a nonintegrable mass
variation \cite{HuCaiWang2026}; a related rotating Ricci-flat geometry also
exhibits thermodynamic quantities whose dependence on the external-field
parameter requires careful interpretation \cite{MaLu2026}.  These observations
suggest that the static solution should be analyzed at the level of the
symplectic structure alongside its mass formulas.

A second clue comes from the global geometry.  Herdeiro and Novo showed that
the nominally vacuum, zero-mass background associated with this solution has a
distributional source on a semi-infinite annular domain in Weyl coordinates
\cite{HerdeiroNovo2026}.  A direct comparison with the black-hole phase-space
flux requires the finite-\(m\) Weyl map, because the map degenerates at the
limiting boundary and its support changes in the zero-mass limit.  The
appropriate dynamical quantity is also a source response, quadratic in
solution-space variations and carrying more information than the static
surface density.

We address three connected problems.  First, we compute the complete
surface-charge one-form for the regular two-parameter family and determine its
integrability for fixed and field-dependent time generators.  Second, we
separate homothetic equivalence from gauge degeneracy by evaluating
exact-symmetry charges along the scaling orbit.  Third, we construct the
finite-mass Weyl boundary and compare its Brown--York canonical response with
the Lee--Wald symplectic flux.

The results give a unified answer.  With the asymptotic coordinate \(t\) held
fixed, the charge one-form is finite on the horizon and at infinity but has a
nonzero curl.  Its integrating factor
\(N=(1+B^2m^2)^{-1/2}\) defines the normalized energy
\(H=m(1+B^2m^2)^{-3/2}\).  The dimensionless combination \(y=Bm\)
nevertheless changes the intrinsic shape of the horizon and of the limiting
Weyl surface at fixed \(H\).  The Brown--York response of that surface equals
the bulk symplectic flux, including its coefficient.  Consequently \(y\) is a
physical source modulus, and variation of this modulus supplies the work term
in the fixed-time Hamiltonian.  This dual role resolves the apparent
competition between hair and source interpretations.

We use \(G=1\), metric signature \((-+++)\), and orientation
\(\epsilon_{tr x\phi}>0\).  The azimuthal coordinate has period \(2\pi\)
after the regularity factor has been incorporated into the metric.

\section{Static type-I solution and thermodynamic data}
\label{sec:geometry}

\subsection{Metric and physical domain}

In coordinates \(x^\mu=(t,r,x,\phi)\), with \(x=\cos\theta\), introduce
\begin{align}
 P&=1+B^2r\left[r+
 \bigl(2m+(B^2m^2-1)r\bigr)x^2\right], \nonumber\\
 w&=\sqrt{P},\qquad
 f=\left(1-\frac{2m}{r}-B^2m^2\right)(1+B^2r^2), \nonumber\\
 u&=1+B^2mrx^2+w,\qquad
 \Delta_\phi=\frac{1}{1+B^2m^2}.
\label{eq:metric-functions}
\end{align}
The regular static type-I line element is
\begin{align}
\dd s^2={}&A\left[-f\,\dd t^2+\frac{\dd r^2}{f}\right.\nonumber\\
&\left.+\frac{r^2\dd x^2}{(1-x^2)(1+B^2m^2x^2)}\right]\nonumber\\
&+\frac{4r^2(1-x^2)(1+B^2m^2x^2)\Delta_\phi^2}{u^2}
\dd\phi^2,\qquad\nonumber\\
A={}&\frac{u^2}{4w^4}.
\label{eq:metric}
\end{align}
The square root is taken on the positive exterior branch.  Our black-hole
domain is
\begin{equation}
 m>0,\qquad 0<y\equiv Bm<1,\qquad r>r_h,
\label{eq:domain}
\end{equation}
with the \(B=0\) Schwarzschild line included by continuity.  The factor
\(\Delta_\phi\) makes both components of the rotation axis regular while
\(\phi\in[0,2\pi]\).  Because \(\Delta_\phi\) depends on the parameters, it
must be varied together with the metric when constructing the regular solution
phase space.

The metric is Ricci-flat and generically Petrov type I
\cite{Astorino2026}.  Its determinant takes the simple form
\begin{equation}
 \sqrt{-g}=\frac{r^2u^2\Delta_\phi}{4w^6},
\label{eq:determinant}
\end{equation}
which is useful for both surface charges and the Weyl transformation.

\subsection{Axis regularity and algebraic character}

The regularity factor in Eq.~\eqref{eq:metric-functions} can be recovered
directly from the geometry.  Near either component of the axis,
\(w=1+B^2mr\), and the ratio between azimuthal circumference and proper radius
obeys
\begin{equation}
 \lim_{x\to\pm1}
 4\pi^2\frac{(\partial_x g_{\phi\phi})^2}
 {4g_{\phi\phi}g_{xx}}=(2\pi)^2 .
\label{eq:axis-regularity}
\end{equation}
The same value is obtained on both axes.  A change of \(m\) or \(B\) changes
the unrescaled angular period, which explains why
\(\Delta_\phi(m,B)\) participates in the tangent variations used in
Sec.~\ref{sec:cps}.  This global regularity condition selects a definite
two-parameter phase space.

The solution is algebraically general for generic \(mB\).  In terms of the
usual complex Weyl invariants \(I\) and \(J\), the discriminant
\(I^3-27J^2\) is nonzero at exterior points and hence on an open dense set.
It vanishes in the correlated \(m=0\) background, which is type D.  The
change in algebraic type accompanies the singular Weyl boundary limit
developed in Sec.~\ref{sec:weyl}.

\subsection{Horizon geometry}

The outer Killing horizon is the zero of the first factor in \(f\),
\begin{equation}
 r_h=\frac{2m}{1-B^2m^2}.
\label{eq:horizon-radius}
\end{equation}
At the horizon the square root simplifies to an angle-independent value,
\begin{equation}
 w_h=\frac{1+B^2m^2}{1-B^2m^2}.
\label{eq:horizon-w}
\end{equation}
This cancellation makes the surface gravity constant.  It also reduces the
horizon area element to
\begin{equation}
 \sqrt{g_{xx}g_{\phi\phi}}\big|_{r_h}
 =\frac{r_h^2\Delta_\phi}{w_h^2}
 =\frac{4m^2}{(1+B^2m^2)^3},
\label{eq:horizon-area-density}
\end{equation}
independent of \(x\).  Integration over
\([-1,1]\times[0,2\pi]\) produces the area below.
The surface gravity of \(\partial_t\), the Hawking temperature, the area, and
the entropy are
\begin{align}
 \kappa&=\frac{(1+B^2m^2)^2}{4m},&
 T&=\frac{(1+B^2m^2)^2}{8\pi m},\nonumber\\
 A_h&=\frac{16\pi m^2}{(1+B^2m^2)^3},&
 S&=\frac{A_h}{4}
 =\frac{4\pi m^2}{(1+B^2m^2)^3}.
\label{eq:horizon-thermo}
\end{align}
These expressions obey the Bekenstein--Hawking and Hawking relations
\cite{Bekenstein1973,Hawking1975,Wald1993}.  A Komar integral associated with
the same Killing field gives
\begin{equation}
 M=\frac{m}{1+B^2m^2},\qquad M=2TS.
\label{eq:komar-smarr}
\end{equation}
The second equality is the static Smarr relation
\cite{Komar1959,Smarr1973,BardeenCarterHawking1973}.
More explicitly, for any closed surface homologous to the horizon,
\begin{equation}
 M_{\mathrm K}[\partial_t]
 =-\frac{1}{8\pi}\int_S \star\dd(\partial_t)^\flat .
\label{eq:komar-integral}
\end{equation}
Vacuum closure makes this integral independent of the chosen finite-radius
surface.  Evaluation at the horizon gives \(2TS\), while the limiting angular
integral gives the first expression in Eq.~\eqref{eq:komar-smarr}.  Their
equality fixes the normalization of \(t\) used in the fixed-generator
analysis.

The deformation is visible in an invariant dimensionless ratio of horizon
circumferences.  Let \(C_e\) be the equatorial circumference and \(C_p\) the
length of a closed meridian.  Direct integration of the induced horizon
metric gives
\begin{align}
 C_e&=\frac{4\pi m}{1+B^2m^2},&
 C_p&=\frac{8m\,\cE(-B^2m^2)}{(1+B^2m^2)^2},\nonumber\\
 \frac{C_p}{C_e}
 &=\frac{2\cE(-y^2)}{\pi(1+y^2)},
\label{eq:shape-ratio}
\end{align}
where \(\cE(q)\) is the complete elliptic integral of the second kind with
parameter \(q\).  The ratio decreases strictly from unity on \(0<y<1\), as
proved in Appendix~\ref{app:proofs}.  Thus \(y\) changes horizon shape even
when an overall length scale is removed.

Equations~\eqref{eq:horizon-thermo} and \eqref{eq:komar-smarr} expose the
thermodynamic tension:
\begin{equation}
 \dd M-T\dd S
 =\frac{Bm^3}{(1+B^2m^2)^2}\,\dd B
 +\frac{B^2m^2}{(1+B^2m^2)^2}\,\dd m .
\label{eq:thermo-tension}
\end{equation}
The right-hand side is generically nonzero.  We now determine which part of
this difference is encoded by the covariant symplectic structure.

\section{Covariant phase-space charges and integrability}
\label{sec:cps}

\subsection{Surface charge}

For the Einstein--Hilbert Lagrangian
\(\bm{L}=R\eps/(16\pi)\), a metric variation
\(h_{\mu\nu}=\delta g_{\mu\nu}\) produces the standard symplectic potential
\cite{LeeWald1990,IyerWald1994}
\begin{equation}
 \ThetaLW_{\alpha\beta\gamma}[g;h]
 =\frac{1}{16\pi}\epsilon_{\mu\alpha\beta\gamma}
 \left(\nabla_\nu h^{\mu\nu}-\nabla^\mu h\right),
\label{eq:theta}
\end{equation}
where \(h=g^{\mu\nu}h_{\mu\nu}\).  The Noether charge two-form is
\begin{equation}
 (\Q_\xi)_{\alpha\beta}
 =-\frac{1}{16\pi}\epsilon_{\mu\nu\alpha\beta}
 \nabla^\mu\xi^\nu .
\label{eq:noether-charge}
\end{equation}
For a generator that may depend on the solution parameters, the adjusted
Iyer--Wald surface charge is
\begin{equation}
 \kform_\xi[g;\delta g]
 =\delta\Q_\xi-\Q_{\delta\xi}-i_\xi\ThetaLW[g;\delta g].
\label{eq:iw-charge}
\end{equation}
The term \(\Q_{\delta\xi}\) separates variation of the generator from the
physical metric variation.  Appendix~\ref{app:cps} gives the independent
Barnich--Brandt component formula and its equality to
\eqref{eq:iw-charge} for this family.

Let \(h_m=\partial_m g\) and \(h_B=\partial_B g\), including the parameter
dependence of \(\Delta_\phi\).  Integrating the \(x\phi\) component of
\eqref{eq:iw-charge} on a surface \(t,r=\mathrm{const}\) defines
\begin{equation}
 \alpha=\int_{S_r}\kform_{\partial_t}
 =C_m\,\dd m+C_B\,\dd B .
\label{eq:charge-one-form-def}
\end{equation}
The result is independent of \(r\), including the limiting surfaces at the
horizon and infinity:
\begin{equation}
 C_m=\frac{1-2B^2m^2}{(1+B^2m^2)^2},
 \qquad
 C_B=-\frac{3Bm^3}{(1+B^2m^2)^2}.
\label{eq:charge-one-form}
\end{equation}
At \(B=0\), \(C_m=1\), fixing the sign and the factor \(16\pi\) against the
Schwarzschild mass.  On the horizon the same coefficients satisfy
\begin{equation}
 C_p^{\,h}=T\,\partial_p S,\qquad p\in\{m,B\}.
\label{eq:horizon-charge}
\end{equation}
The equality between horizon and infinity follows locally from
\(\dd\kform_{\partial_t}=0\) for tangent solutions, together with vanishing
flux at the two regular axis endpoints.  It is therefore insensitive to the
coordinate degeneration of the limiting Weyl image.

It is useful to distinguish \(\alpha\) from the exterior derivative of the
Komar expression \(M(m,B)\).  A Komar integral evaluates the Noether charge
of \(\partial_t\) on each individual solution.  A Hamiltonian variation also
contains \(-i_{\partial_t}\ThetaLW\), which compares neighboring solutions and
is sensitive to flux through the boundary.  Their difference is
Eq.~\eqref{eq:thermo-tension}:
\begin{equation}
 \dd M-\alpha
 =\frac{B^2m^2\,\dd m+Bm^3\,\dd B}
 {(1+B^2m^2)^2}.
\label{eq:komar-versus-cps}
\end{equation}
Thus the value of the Komar integral and the integrability of the covariant
Hamiltonian answer distinct questions.  The Smarr identity concerns the
former on a fixed solution; the first law concerns the latter over solution
space.  Their coexistence is typical when boundary data vary.

\subsection{Nonintegrability and symplectic flux}

The exterior derivative of \(\alpha\) on the two-dimensional solution space is
\begin{align}
 \dd\alpha
 &=
 \left(\partial_m C_B-\partial_B C_m\right)
 \dd m\wedge\dd B\nonumber\\
 &=-\frac{Bm^2}{(1+B^2m^2)^2}\,\dd m\wedge\dd B .
\label{eq:charge-curl}
\end{align}
Hence the fixed-time charge fails the Hamiltonian integrability condition on
the full regular family.  Integrating \(\alpha\) along two paths with the same endpoints
gives a difference equal to the integral of \(\dd\alpha\) over the enclosed
parameter-space rectangle.

For illustration, choose the rectangle bounded by
\((m,B)=(1,1/5),(2,1/5),(1,1/3),(2,1/3)\), which lies in the
physical domain.  The difference between paths that vary \(m\) first and
\(B\) first is
\begin{align}
 \Delta_\square H={}&-\frac52\arctan\frac25
 -\frac32\arctan\frac13\nonumber\\
 &+\frac52\arctan\frac15
 +\frac32\arctan\frac23\neq0 .
\label{eq:path-difference}
\end{align}
This finite loop calculation reaches the same conclusion as the local curl.
Selecting one of the paths would amount to supplying additional boundary
information.

The Lee--Wald current,
\begin{equation}
 \omegaf[g;h_1,h_2]
 =\delta_1\ThetaLW[g;h_2]-\delta_2\ThetaLW[g;h_1],
\label{eq:lee-wald-current}
\end{equation}
identifies the spacetime origin of the obstruction.  With the orientation
fixed above,
\begin{equation}
 \dd\alpha
 =-\int_{S_r}i_{\partial_t}\omegaf[g;h_m,h_B]\,
 \dd m\wedge\dd B .
\label{eq:curl-flux}
\end{equation}
The integrand equality holds pointwise after the exact field equations and
linearized field equations are used.  Its integral is independent of finite
\(r\), and both the direct \(r\rightarrow\infty\) integral and the integral of
the limiting angular density give Eq.~\eqref{eq:charge-curl}.  The convergence
near \(x=\pm1\) is slower than at the equator, reflecting the degeneration of
the asymptotic Weyl map, while the integrated limit remains finite.

\subsection{Field-dependent time normalization}

Consider the normalized exact symmetry
\begin{equation}
 \bar{\xi}=N\partial_t,\qquad
 N=\frac{1}{\sqrt{1+B^2m^2}} .
\label{eq:barred-generator}
\end{equation}
Since \(N\) is constant on spacetime and varies over solution space,
\(\delta\bar{\xi}=(\delta N)\partial_t\).  Substitution into
\eqref{eq:iw-charge}, including \(-\Q_{\delta\bar{\xi}}\), gives
\begin{equation}
 \kform_{\bar{\xi}}=N\kform_{\partial_t},
 \qquad
 \bar{\alpha}=N\alpha
 =\dd\!\left[\frac{m}{(1+B^2m^2)^{3/2}}\right].
\label{eq:barred-charge}
\end{equation}
We denote the finite charge by
\begin{equation}
 H=\frac{m}{(1+B^2m^2)^{3/2}},\qquad H(m,0)=m .
\label{eq:H-def}
\end{equation}
The temperature rescales linearly with the generator.  In the normalized
frame,
\begin{align}
 \bar M&=H,\qquad
 \bar T=NT=\frac{1}{8\pi H},\nonumber\\
 S&=4\pi H^2,\qquad
 \dd\bar M=\bar T\,\dd S .
\label{eq:barred-thermodynamics}
\end{align}
All three thermodynamic functions depend only on the scale coordinate \(H\).
The remaining coordinate \(y\) is invisible to this first law and visible to
dimensionless geometry, a distinction developed in the next section.
The subtraction in Eq.~\eqref{eq:iw-charge} is essential: retaining the
variation of \(N\) inside \(\delta\Q_{\bar\xi}\) without subtracting
\(\Q_{\delta\bar\xi}\) would assign a spurious physical charge to the change
of generator.

Table~\ref{tab:generators} summarizes the three exact symmetries used below.
The entropy generator is the horizon Killing field normalized by inverse
temperature, as in the Noether-charge derivation of black-hole entropy
\cite{Wald1993,IyerWald1994}.

\begin{table}[t]
\caption{\label{tab:generators}Exact-symmetry generators and their charge
variations on the regular solution space.}
\begin{ruledtabular}
\begin{tabular}{lccc}
generator & charge one-form & integrable & finite charge\\
\(\partial_t\) & \(\alpha\) & no & path dependent\\
\(N\partial_t\) & \(N\alpha\) & yes & \(H\)\\
\(T^{-1}\partial_t\) & \(T^{-1}\alpha\) & yes & \(S\)\\
\end{tabular}
\end{ruledtabular}
\end{table}

\subsection{The role of the angular ensemble}

A useful comparison holds the numerical value of \(\Delta_\phi\) fixed while
\(m\) and \(B\) vary.  Such variations change the conical deficit on the axis.
They remain solutions of the vacuum equations away from the axis and isolate
the contribution associated with regularizing the azimuthal orbit.  The
surface-charge coefficients in this comparison are
\begin{align}
 C_m^{(\Delta)}
 &=-\Delta_\phi\frac{(Bm-1)(Bm+1)}{1+B^2m^2},
 \nonumber\\
 C_B^{(\Delta)}
 &=-\frac{2B\Delta_\phi m^3}{1+B^2m^2},
\label{eq:fixed-period-coefficients}
\end{align}
and their curl is
\begin{equation}
 \dd\alpha_\Delta
 =-\frac{2B\Delta_\phi m^2}{1+B^2m^2}
 \,\dd m\wedge\dd B .
\label{eq:fixed-period-curl}
\end{equation}
The coefficient differs from Eq.~\eqref{eq:charge-curl}.  The regular ensemble
is selected by Eq.~\eqref{eq:axis-regularity}; the comparison displays the
part of the obstruction that would be accompanied by an axis source.

\subsection{Hamiltonian meaning of the obstruction}

Integrability is a statement about the pair consisting of a generator and a
phase space.  The fixed vector \(\partial_t\) supplies a common comparison of
time translations across the family, and Eq.~\eqref{eq:charge-curl} says that
this comparison admits symplectic flux when \(m\) and \(B\) are varied
independently.  The equality in Eq.~\eqref{eq:curl-flux} rules out a mismatch
between a horizon formula and a separate asymptotic prescription: the same
two-form is measured on every homologous surface.

One possible treatment of flux at a boundary is to modify the symplectic
potential by boundary data, as in the general Wald--Zoupas framework
\cite{WaldZoupas2000}.  For the present static family, keeping the standard
Einstein--Hilbert representative has two advantages.  It preserves the direct
Iyer--Wald/Barnich--Brandt equivalence, and it leaves the coefficient of the
obstruction available for comparison with the independently constructed Weyl
source.  Section~\ref{sec:weyl} shows that this coefficient is exactly the
canonical response of the limiting boundary.  A boundary polarization that
fixes the source would remove the corresponding variation from the phase
space; allowing the source to vary produces the work term derived in
Sec.~\ref{sec:work}.

The integrating factor \(N\) provides a second Hamiltonian description.  It
changes which stationary vector is compared between neighboring solutions and
therefore changes the energy polarization.  Its role is analogous to choosing
a properly normalized frame in other non-asymptotically flat black-hole
families.  The simultaneous existence of a flux-carrying fixed-time
description and a flux-free normalized description is a physical feature of
field-dependent exact symmetries, fully accounted for by
\(-\Q_{\delta\xi}\) in Eq.~\eqref{eq:iw-charge}.

\section{Scaling, solution-space coordinates, and exact symmetries}
\label{sec:scaling}

\subsection{Scale and shape coordinates}

The normalized charge suggests coordinates adapted to scale and shape:
\begin{equation}
 y=Bm,\qquad
 H=\frac{m}{(1+y^2)^{3/2}}.
\label{eq:Hy-map}
\end{equation}
Their inverse and Jacobian are
\begin{align}
 m&=H(1+y^2)^{3/2},&
 B&=\frac{y}{H(1+y^2)^{3/2}},\nonumber\\
 \det\frac{\partial(H,y)}{\partial(m,B)}&=H.
\label{eq:Hy-inverse}
\end{align}
The map is regular throughout Eq.~\eqref{eq:domain}.  In these coordinates the
fixed-time charge becomes especially transparent:
\begin{equation}
 \alpha=\sqrt{1+y^2}\,\dd H,\qquad
 \dd\alpha=-\frac{y}{\sqrt{1+y^2}}\,\dd H\wedge\dd y .
\label{eq:alpha-Hy}
\end{equation}
The scale \(H\) is the integrable energy of \(\bar\xi\), while \(y\) is the
dimensionless deformation that carries the symplectic obstruction.

The complete local classification of integrating factors follows immediately.
For a generator \(\xi_\mu=\mu(H,y)\partial_t\), closure of
\(\mu\alpha\) requires
\(\partial_y[\mu\sqrt{1+y^2}]=0\).  Therefore
\begin{equation}
 \mu=N F(H),\qquad N=(1+y^2)^{-1/2},
\label{eq:integrating-factors}
\end{equation}
where \(F\) is an arbitrary local function.  Requiring
\(\mu(m,0)=1\) along the entire Schwarzschild line sets \(F(H)=1\).
Thus the normalization \eqref{eq:barred-generator} is unique in the class
whose energy agrees with Schwarzschild mass for every \(m>0\).  The entropy
generator belongs to the same family:
\begin{equation}
 \mu_S=\frac{1}{T}=N(8\pi H),\qquad
 \mu_S\alpha=\dd S,\qquad S=4\pi H^2.
\label{eq:entropy-generator}
\end{equation}
The solution phase-space method for exact symmetries
\cite{HajianSheikhJabbari2016} yields the same three integrability statements
in Table~\ref{tab:generators}.

\subsection{Homothety and parameter count}

For \(B>0\), the dimensionful parameter can be extracted from the metric by
the coordinate dilation
\[
 T=Bt,\qquad R=Br.
\]
Writing \(\Phi_B:(t,r,x,\phi)\mapsto(T,R,x,\phi)\), one finds the exact
homothety
\begin{equation}
 g(m,B)=B^{-2}\Phi_B^*g(y,1).
\label{eq:homothety}
\end{equation}
Consequently, geometries modulo coordinate dilation and a choice of overall
unit scale form a one-dimensional shape family labeled by \(y\).

The tangent to the scaling orbit is
\begin{align}
 V&=B\partial_B-m\partial_m,\nonumber\\
 Bh_B-mh_m&=\Lie_D g-2g,\qquad
 D=t\partial_t+r\partial_r .
\label{eq:scaling-variation}
\end{align}
The constant Weyl term in Eq.~\eqref{eq:scaling-variation} is significant at
fixed \(G\): the Einstein--Hilbert density has nonzero Weyl weight, so this
homothety defines a physical phase-space direction.  Exact-symmetry charges
give a direct phase-space test,
\begin{align}
 V(y)&=0,\qquad V(H)=-H,\nonumber\\
 \alpha(V)&=-M,\qquad \dd S(V)=-2S.
\label{eq:scaling-charges}
\end{align}
The scaling direction therefore lies outside the presymplectic kernel.  The
fixed-\(G\) physical solution space remains two-dimensional.  A simultaneous
rescaling of the numerical value assigned to \(G\) implements a change of
units and leads to the one-dimensional homothety quotient described by
\eqref{eq:homothety}.

There is also a deformation tangent to \(H=\mathrm{const}\).  It has vanishing
barred energy and entropy variation, yet Eq.~\eqref{eq:shape-ratio} changes
strictly with \(y\).  This gives an intrinsic distinction between a
thermodynamically null direction for one generator and a diffeomorphism
degeneracy.

\section{Weyl boundary source and symplectic flux}
\label{sec:weyl}

\subsection{Canonical Weyl map}

Static axisymmetric vacuum metrics admit Weyl coordinates \((\rho,z)\)
\cite{Weyl1917,BachWeyl1922}.  Define
\begin{equation}
 F=r^2f,\qquad
 X=(1-x^2)(1+B^2m^2x^2).
\label{eq:FX-def}
\end{equation}
The canonical radius follows directly from the two Killing directions,
\(\rho^2=-g_{tt}g_{\phi\phi}\).  Its harmonic conjugate can be chosen so that
\begin{align}
 \rho^2&=\frac{\Delta_\phi^2FX}{P^2},\nonumber\\
 z&=\frac{\Delta_\phi x(r-m)(1+B^2mr)}{P}.
\label{eq:weyl-map}
\end{align}
These expressions satisfy the Cauchy--Riemann system associated with the
two-dimensional orbit metric.  The Jacobian degenerates at the limiting
surface \(r=\infty\), which is precisely why its image must be obtained before
performing the massless limit.

\begin{figure*}[t]
\includegraphics[width=\textwidth]{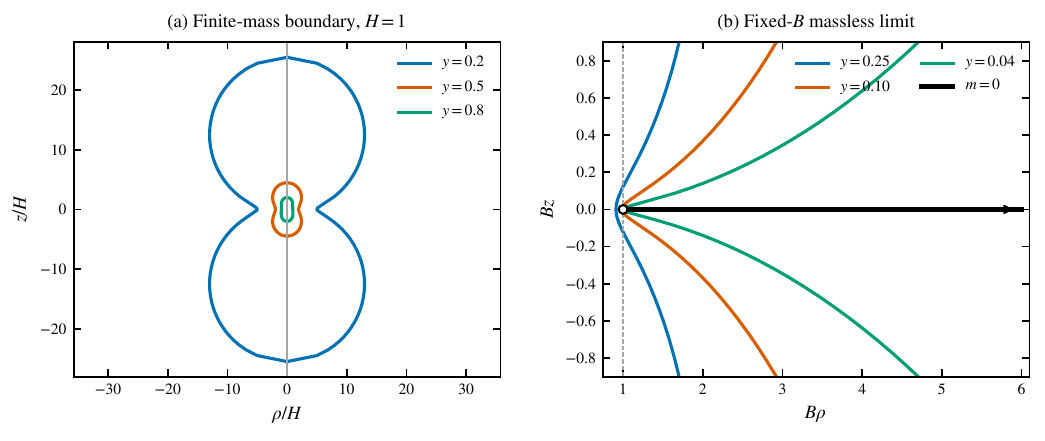}
\caption{\label{fig:weyl}Weyl image of the limiting boundary.\ (a) Meridional
profiles at fixed \(H=1\); the reflected \(\rho<0\) branch visualizes the
closed surface of revolution. Increasing \(y\) changes its dimensionless
shape.\ (b) At fixed \(B=1\), the finite-mass profiles approach the plane
\(z=0\) nonuniformly and open into the semi-infinite domain
\(\rho\geq1/B\).}
\end{figure*}

At fixed material coordinate \(x\), the image of spatial infinity is
\begin{align}
 \rho_\infty(x)
 &=\frac{H\sqrt{1+y^2}}{y}\,
 \frac{\sqrt{(1-y^2)(1-x^2)(1+y^2x^2)}}{A_y},
 \nonumber\\
 z_\infty(x)
 &=\frac{H\sqrt{1+y^2}\,x}{A_y},
 \qquad
 A_y=1-(1-y^2)x^2 .
\label{eq:weyl-boundary}
\end{align}
For \(0<y<1\), \(\rho_\infty>0\) on \(-1<x<1\), both endpoints lie on
the axis, and
\begin{equation}
 \frac{\dd z_\infty}{\dd x}
 =H\sqrt{1+y^2}\,
 \frac{1+(1-y^2)x^2}{A_y^2}>0 .
\label{eq:weyl-monotonic}
\end{equation}
The meridional curve is consequently embedded; revolving it about the axis
produces a closed boundary surface.  Figure~\ref{fig:weyl} displays the
surface profiles for three values of \(y\).

The equatorial radius and the north-axis height provide two simple geometric
scales,
\begin{equation}
 \rho_{\mathrm{eq}}=\frac{H\sqrt{1-y^4}}{y},
 \qquad
 z_{\mathrm N}=\frac{H\sqrt{1+y^2}}{y^2}.
\label{eq:weyl-scales-main}
\end{equation}
Their ratio \(z_{\mathrm N}/\rho_{\mathrm{eq}}
=[y\sqrt{1-y^2}]^{-1}\) depends only on \(y\).  The limiting boundary
therefore carries the same scale-shape separation already seen at the
horizon, although the two shape functions are different.

The zero-mass background is reached by \(m\rightarrow0\) at fixed \(B\), so
both \(H\) and \(y\) tend to zero in a correlated way.  For fixed
\(-1<x<1\), Eq.~\eqref{eq:weyl-boundary} becomes
\begin{equation}
 z_\infty=0,\qquad
 \rho_\infty=\frac{1}{B\sqrt{1-x^2}}\geq\frac{1}{B}.
\label{eq:massless-annulus}
\end{equation}
Taking the axis endpoints and the zero-mass limit in opposite orders yields
different results.  The closed finite-mass surface opens into the
semi-infinite annular domain found in
Ref.~\cite{HerdeiroNovo2026}.  Similar nonuniform zero-mass limits are known
to alter global extensions even when local curvature limits appear smooth
\cite{GibbonsVolkov2017}.

\subsection{Static source and canonical response}

For the massless background, the natural reflected extension across
\(z=0\) has surface density
\begin{equation}
 \sigma(\rho)=
 \frac{B}{2\pi\sqrt{B^2\rho^2-1}},
 \qquad \rho>\frac{1}{B}.
\label{eq:HN-density}
\end{equation}
It diverges integrably at the inner edge and falls as \(1/\rho\) at large
radius \cite{HerdeiroNovo2026}.  The seam-oriented curvature and the
corresponding mixed Israel stress are, in the \((t,\rho,\phi)\) ordering,
\begin{equation}
 \widehat K^a{}_b=(2B,2B,-2B),\qquad
 S^a{}_b=(0,0,B/\pi)
\label{eq:HN-israel}
\end{equation}
on the massless sheet, with the proper-normal delta function converted to the
Weyl \(\delta(z)\) in the distributional curvature.  In the four-dimensional
ordering \((t,\rho,z,\phi)\), the singular Ricci tensor is
\begin{equation}
 \left.R^\mu{}_\nu\right|_{\mathrm{dist}}
 =8B\sqrt{B^2\rho^2-1}\,\delta(z)\,
 \operatorname{diag}(-1,-1,-1,1).
\label{eq:HN-ricci}
\end{equation}
Equations~\eqref{eq:HN-density}, \eqref{eq:HN-israel}, and
\eqref{eq:HN-ricci} agree after the proper-distance and Weyl-coordinate
measures are related at the sheet.

For finite mass, take timelike surfaces \(\Sigma_R:r=R\) with induced metric
\(h_{ab}\).  Let \(n^\mu_{\rm out}\) point toward increasing \(r\), and define
\(K^{\rm out}_{ab}=h_a{}^\mu h_b{}^\nu\nabla_\mu n^{\rm out}_\nu\) and
\(K_{\rm out}=h^{ab}K^{\rm out}_{ab}\).  The one-sided canonical momentum is
\begin{equation}
 \Pi^{ab}
 =\frac{\sqrt{-h}}{16\pi}
 \left(K_{\rm out}^{ab}-K_{\rm out}h^{ab}\right).
\label{eq:BY-momentum}
\end{equation}
It is the radial canonical datum conjugate to \(h_{ab}\)
\cite{BrownYork1993,Poisson2004}.  For the reflected completion, the normal
from the seam into either retained sheet is
\(\widehat n^\mu=-n^\mu_{\rm out}\), so that
\(\widehat K_{ab}=-K^{\rm out}_{ab}\).  The Israel tensor is
\begin{equation}
 S_{ab}
 =-\frac{1}{4\pi}
 \left(\widehat K_{ab}-\widehat K h_{ab}\right)
 =\frac{1}{4\pi}
 \left(K^{\rm out}_{ab}-K_{\rm out}h_{ab}\right),
\label{eq:Israel-stress}
\end{equation}
where reflection gives a jump of twice the seam-oriented curvature
\cite{Israel1966}.  This reflected extension supplies a concrete matter
interpretation.  The Weyl map and the one-sided pair \((h_{ab},\Pi^{ab})\)
already determine the symplectic response.

Transform the tangent fields from \((m,B)\) to \((H,y)\) and keep the material
coordinate \(x\) fixed on \(\Sigma_R\).  The antisymmetrized boundary
canonical response is
\begin{equation}
 \mathcal{W}_{Hy}
 =\delta_H\Pi^{ab}\,\delta_y h_{ab}
 -\delta_y\Pi^{ab}\,\delta_H h_{ab}.
\label{eq:boundary-response}
\end{equation}
The radial Hamiltonian identity gives the pointwise equality
\begin{equation}
 \lim_{R\to\infty}\mathcal{W}_{Hy}\,\dd t\wedge\dd x\wedge\dd\phi
 =i_{\partial_t}\omegaf[g;h_H,h_y].
\label{eq:BY-LW-match}
\end{equation}
Both sides use the same Einstein--Hilbert representative, with no boundary
counterterm.  Integrating a constant-\(t\) section yields
\begin{equation}
 \int_{\partial\Sigma}i_{\partial_t}\omegaf[g;h_H,h_y]
 =\frac{y}{\sqrt{1+y^2}},
\label{eq:integrated-flux}
\end{equation}
which reproduces Eq.~\eqref{eq:alpha-Hy} through
\eqref{eq:curl-flux}.

The Weyl pushforward of Eq.~\eqref{eq:boundary-response} contains the
parameter-dependent displacement
\((\delta\rho_\infty,\delta z_\infty)\).  At fixed Weyl coordinates this
produces derivative-of-delta terms in addition to the variation of the local
stress.  Keeping \(x\) fixed is the Lagrangian description of the moving
source; holding \((\rho,z)\) fixed gives the equivalent Eulerian
description.  Appendix~\ref{app:boundary} records the transformation and the
coefficient match.

This distinction clarifies the relation between \(\sigma\) and the
symplectic flux.  The density \eqref{eq:HN-density} is a one-point property of
the massless background.  By contrast, \(\mathcal{W}_{Hy}\) is a two-form on
solution space and probes the antisymmetrized variation of the boundary
canonical potential.  Its ingredients are the variation of the stress, the
variation of the induced metric, and the displacement of the embedded
support.  Equation~\eqref{eq:BY-LW-match} compares objects with the same
variational degree and retains all three contributions.

The result also has a radial-Hamiltonian interpretation.  On shell, the
Einstein--Hilbert symplectic current through a timelike radial boundary is the
field-space exterior derivative of
\(\Pi^{ab}\delta h_{ab}\).  The equality holds at finite \(R\), before the
limiting map to Weyl coordinates.  Taking \(R\rightarrow\infty\) then trades a
regular coordinate surface for the closed embedding
\eqref{eq:weyl-boundary}.  The fixed-\(x\) description keeps this operation
regular, and the displacement terms give its distributional Weyl
representation.

The normalized generator \(N\partial_t\) has zero integrated flux, since its
charge is \(\dd H\).  The source geometry continues to vary with \(y\) at
fixed \(H\), as Fig.~\ref{fig:weyl} (a) makes explicit.  Time normalization
changes the Hamiltonian polarization and leaves the geometric source modulus
intact.

\section{External work and black-hole mechanics}
\label{sec:work}

The scale-shape coordinates turn the thermodynamics into a standard
source-response form.  Equations~\eqref{eq:horizon-thermo},
\eqref{eq:komar-smarr}, and \eqref{eq:Hy-inverse} become
\begin{equation}
 M=H\sqrt{1+y^2},\qquad
 S=4\pi H^2,\qquad
 T=\frac{\sqrt{1+y^2}}{8\pi H}.
\label{eq:thermo-Hy}
\end{equation}
Taking the exterior derivative gives the global first law
\begin{equation}
 \dd M=T\dd S+\Psi_y\dd y,\qquad
 \Psi_y=\frac{Hy}{\sqrt{1+y^2}},
\qquad M=2TS .
\label{eq:global-first-law}
\end{equation}
The source label \(y\) has zero scaling weight, and the Euler relation
therefore contains only \(M\), \(T\), and \(S\).  Its conjugate \(\Psi_y\) is finite over the full
domain \(0<y<1\).

One may instead use \(B\) as a local source coordinate at fixed entropy.  Since
fixed \(S\) is equivalent to fixed \(H\),
\begin{equation}
 BH=\frac{y}{(1+y^2)^{3/2}},\qquad
 \left(\frac{\partial B}{\partial y}\right)_H
 =\frac{1-2y^2}{H(1+y^2)^{5/2}}.
\label{eq:B-turning-map}
\end{equation}
The map has a maximum at
\begin{equation}
 y_*=\frac{1}{\sqrt2},\qquad
 B_{\max}=\frac{2}{3\sqrt3\,H}.
\label{eq:B-turning-point}
\end{equation}
Away from this turning point there are two local \(B\) branches, and
\begin{align}
 \dd M&=T\dd S+\Phi_B\dd B,\nonumber\\
 \Phi_B&=\Psi_y\left(\frac{\partial y}{\partial B}\right)_H
 =\frac{H^2y(1+y^2)^2}{1-2y^2}\nonumber\\
 &=\frac{Bm^3}{(1+y^2)(1-2y^2)} .
\label{eq:B-work}
\end{align}
The divergence of \(\Phi_B\) at \(y_*\) is the inverse-Jacobian singularity of
the \((S,B)\) chart.  The invariant work one-form remains regular:
\begin{equation}
 \left.\Psi_y\dd y\right|_{y_*}
 =\frac{H}{\sqrt3}\,\dd y .
\label{eq:turning-regular}
\end{equation}
In the convention commonly used for magnetized black holes,
\(\dd M=T\dd S-\mu\dd B\), the magnetic-response symbol is
\(\mu=-\Phi_B\) \cite{GibbonsPangPope2014}.  Here the origin of the source is
gravitational after the solution-generating map, but the thermodynamic
structure is the same: a boundary datum varies and performs work on the
black-hole Hamiltonian.

\subsection{Global and local source coordinates}

The variable \(y\) is globally adapted to the solution space because it is
dimensionless and monotonic along the shape direction at fixed \(H\).
Accordingly, \(\Psi_y\dd y\) remains a regular one-form across the whole
domain.  The variable \(B\) has the conventional interpretation of an
external-field strength and is convenient for comparison with magnetized
black-hole mechanics \cite{GibbonsPangPope2014}.  Its use at fixed entropy is
local.  The lower branch covers \(0<BH<2/(3\sqrt3)\), whereas the upper
branch covers \(1/(2\sqrt2)<BH<2/(3\sqrt3)\).  A common value of \(B\)
therefore corresponds to two physical values of \(y\) precisely when
\(1/(2\sqrt2)<BH<2/(3\sqrt3)\).

\begin{figure}[t]
\includegraphics[width=\columnwidth]{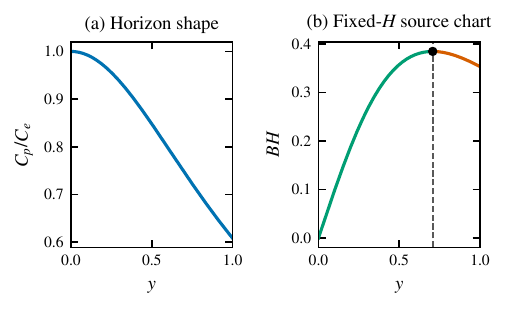}
\caption{\label{fig:shape-work}\ (a) The horizon ratio \(C_p/C_e\) decreases
strictly with the dimensionless source \(y\).\ (b) At fixed \(H\), the map
\(BH=y/(1+y^2)^{3/2}\) has two local branches separated by
\(y=1/\sqrt2\).  The vertical line marks the coordinate turning point.}
\end{figure}

This two-branch structure has a direct geometric reading.  Along the lower
branch \(0<y<1/\sqrt2\), increasing \(B\) at fixed \(H\) increases the
dimensionless deformation.  Along the upper branch
\(1/\sqrt2<y<1\), increasing \(B\) moves toward smaller \(y\).  The sign
change in \(\Phi_B\) follows from this reversal of the coordinate map; the
global response \(\Psi_y\) remains positive.  Figure~\ref{fig:shape-work}
shows both branches together with the strictly monotonic horizon observable.

The Smarr relation contains no explicit source term because \(y\) carries
zero length dimension.  Under \(H\mapsto\lambda H\) at fixed \(y\),
\(M\mapsto\lambda M\), \(S\mapsto\lambda^2S\), and
\(T\mapsto\lambda^{-1}T\).  Euler's theorem then gives \(M=2TS\).
The differential first law probes an independent displacement in \(y\) and
therefore retains \(\Psi_y\dd y\).  This is the same distinction between
Euler scaling and external-source variation that appears in magnetized
Kerr--Newman mechanics, although the present solution is vacuum.

\section{Physical interpretation and conclusions}
\label{sec:conclusions}

\subsection{Geometric modulus and Hamiltonian source}

The two roles of the parameter can now be stated in invariant terms.  A
geometric modulus is a solution-space coordinate whose variation changes a
dimensionless observable and survives quotienting by diffeomorphisms.  At
fixed \(H\), varying \(y=Bm\) changes both
\(C_p/C_e\) and \(z_{\mathrm N}/\rho_{\mathrm{eq}}\).  The first observable is
intrinsic to the horizon; the second belongs to the limiting orbit-space
geometry.  Their dependence on \(y\) rules out an interpretation based solely
on a radial or temporal coordinate rescaling.  The homothetic identity
\eqref{eq:homothety} removes the overall unit scale, while its tangent carries
the nonzero charges \eqref{eq:scaling-charges}.  The physical phase space at
fixed \(G\) therefore has the two coordinates \((H,y)\).

A Hamiltonian source is identified by the response of the canonical
generator.  For the common boundary-time vector \(\partial_t\), variation in
\(y\) produces the curl \eqref{eq:alpha-Hy}.  The obstruction is a genuine
two-form: it is recovered from the Lee--Wald current, from finite loops in
parameter space, and from the Brown--York boundary response.  The global
mechanical relation \eqref{eq:global-first-law} then assigns the conjugate
\(\Psi_y\) to this source coordinate.  These criteria are independent of the
language used for the parameter in the original solution-generating map.

Geometric hair and Hamiltonian source data are compatible classifications.
The former concerns distinguishability of solutions after gauge
identifications.  The latter concerns which data are allowed to vary at the
boundary of a chosen Hamiltonian problem.  An external tidal field, boundary
metric, or electromagnetic field can change intrinsic black-hole geometry
without becoming a Gauss-law charge of the black hole.  The static type-I
family realizes the same pattern entirely within vacuum gravity: the
solution-generating map has transferred the external-field information into
the global geometry.

The normalized vector \(\bar\xi=N\partial_t\) supplies a complementary
polarization.  Its energy \(H\) and temperature \(\bar T\) satisfy the
one-variable first law~\eqref{eq:barred-thermodynamics}.  Schwarzschild
normalization along the complete \(B=0\) family selects \(N\) from the local
class \(\mu=NF(H)\).  The \(y\) direction is thermodynamically null for this
generator and remains geometrically nontrivial.  This illustrates a general
lesson for nonstandard asymptotics: integrability can depend on the
field-dependent normalization of an exact symmetry, whereas the dimension of
the physical solution space is determined by the full presymplectic and
geometric data.

\subsection{Boundary completion}

The Weyl construction locates the source response.  At finite mass, the image
of spatial infinity is a closed surface of revolution.  Its one-sided induced
metric and canonical momentum suffice to calculate the response
\(\mathcal{W}_{Hy}\).  Equation~\eqref{eq:BY-LW-match} shows that this response
equals the bulk symplectic current pointwise, so the boundary accounts for the
entire charge obstruction.  The equality precedes any choice of a
two-sided completion.

The natural reflection identifies the two sides and converts the canonical
momentum into the Israel stress~\eqref{eq:full-israel-components}.  This
completion gives a concrete distributional source for the finite-mass
geometry.  Other global completions can impose different matter
interpretations while retaining the same one-sided canonical data.  The
coefficient match, the closed support, and its motion in solution space are
already fixed by the vacuum exterior.

The relation to the massless background is correspondingly sharp.  Along the
fixed-\(B\) path, the closed surface opens nonuniformly into the plane
\(z=0\), its equator becomes the inner edge \(\rho=1/B\), and the axis
endpoints recede to the noncompact part of the annulus.  The limiting Israel
data reproduce the distributional curvature~\eqref{eq:HN-ricci} and the
density~\eqref{eq:HN-density}.  The finite-mass closed source and the
semi-infinite annulus are two regimes of one boundary geometry, connected by
the boundary-layer expansion~\eqref{eq:massless-boundary-expansion}.

\subsection{Outlook}

The static calculation establishes a sequence of mutually consistent
relations:
\begin{align*}
 \text{surface-charge curl}
 &= -\text{Lee--Wald flux}\\
 &= -\text{boundary canonical response},
\end{align*}
together with the source first law
\(\dd M=T\dd S+\Psi_y\dd y\).  It resolves the title question by assigning
precise meanings to both terms: \(y\) is geometric hair sourced at the
boundary, and \(\Psi_y\dd y\) is its work in the fixed-time Hamiltonian.

Rotation adds an angular generator, a frame choice, and a second conserved
quantity.  The Kerr--Bertotti--Robinson charge obstruction
\cite{HuCaiWang2026} and the Ricci-flat rotating extension
\cite{MaLu2026} make that setting a direct test of the present mechanism.
The decisive calculation will be the matrix of mass and angular-momentum
curls and its comparison with the canonical response of the rotating
orbit-space boundary.  The static result supplies the normalization,
source-coordinate, and zero-mass limits needed for that analysis.

\appendix

\section{Covariant phase-space conventions}
\label{app:cps}

\subsection{Tangent fields}

The only algebraic extension needed to differentiate the metric is generated
by \(w\), with defining relation \(w^2=P\).  Parameter derivatives preserve
this relation when
\begin{equation}
 \partial_p w=\frac{\partial_p P}{2w},\qquad p=m,B,
\label{eq:w-parameter-derivative}
\end{equation}
so that \(\partial_p(w^2-P)=0\) identically.  The fields
\(h_p=\partial_p g\) are therefore exact tangent vectors to the same branch of
the solution.  Differentiating \(G_{\mu\nu}[g(m,B)]=0\) gives
\begin{equation}
 G_{\mu\nu}^{(1)}[h_p]=0,\qquad p=m,B.
\label{eq:linearized-vacuum}
\end{equation}
This observation permits every surface-charge identity below to be reduced
within the quadratic algebra generated by \(w\), without introducing further
radicals.

\subsection{Barnich--Brandt two-form}

For an exact Killing field \(\xi\), the Einstein-gravity
Barnich--Brandt surface density can be written
\cite{BarnichBrandt2002,BarnichCompere2008}
\begin{widetext}
\begin{align}
 k_{\mathrm{BB}}^{\mu\nu}[h;g,\xi]
 =\frac{\sqrt{-g}}{16\pi}\biggl\{&
 \xi^\nu\nabla^\mu h-\xi^\nu\nabla_\rho h^{\mu\rho}
 +\xi_\rho\nabla^\nu h^{\mu\rho}\nonumber\\
 &+\frac12h\nabla^\nu\xi^\mu
 +\frac12h^{\nu\rho}
 \left(\nabla^\mu\xi_\rho-\nabla_\rho\xi^\mu\right)
 -(\mu\leftrightarrow\nu)\biggr\}.
\label{eq:BB-components}
\end{align}
\end{widetext}
Our two-form convention is
\(\kform_{\mathrm{BB}}=\frac12
k_{\mathrm{BB}}^{\mu\nu}(\dd^{\,2}x)_{\mu\nu}\), with
\((\dd^{\,2}x)_{\mu\nu}
=\frac12\epsilon_{\mu\nu\alpha\beta}\dd x^\alpha\wedge\dd x^\beta\).
For the static diagonal metric, only the \(tr\) component contributes on a
constant-\(t,r\) surface.

Expanding Eqs.~\eqref{eq:theta}--\eqref{eq:iw-charge}, using
\(\Lie_\xi g=0\), and collecting covariant derivatives gives
\begin{equation}
 \kform_{\mathrm{IW}}[h;g,\xi]
 =\kform_{\mathrm{BB}}[h;g,\xi]
\label{eq:IW-BB-equivalence}
\end{equation}
for \(h=h_m,h_B\).  Terms proportional to the linearized Einstein tensor
vanish because both variations are tangent to the Ricci-flat solution family.
The equality holds before angular integration.  At \(B=0\), its remaining
density integrates to \(\delta m\), which fixes the orientation and
normalization.

\subsection{Limiting charge densities}

The angular densities provide a local check on the integrated
coefficients~\eqref{eq:charge-one-form}.  Define
\begin{equation}
 q(x,y)=\sqrt{1-(1-y^2)x^2}.
\label{eq:q-def}
\end{equation}
Taking \(r\rightarrow\infty\) at fixed \(-1<x<1\), the coefficient of
\(\dd x\wedge\dd\phi\) for the \(m\) variation is
\begin{align}
 k_\infty^{(m)}={}&
 \frac{y}{4\pi(1+y^2)^2q^4}
 \bigl[y^3x^2-2y^2q-yx^2-y+2q\bigr].
\label{eq:kinf-m}
\end{align}
For the \(B\) variation it is
\begin{align}
 k_\infty^{(B)}={}&
 \frac{1}{4\pi B^2(1+y^2)^2q^4}
 \bigl[
 2y^5x^2-3y^4q-y^3x^2-2y^3
 \nonumber\\
 &\hspace{27mm}
 +2y^2q-yx^2-y+q
 \bigr].
\label{eq:kinf-B}
\end{align}
Each density is even in \(x\) and finite at the endpoints for fixed
\(0<y<1\).  The factors in Eq.~\eqref{eq:kinf-B} cancel sufficiently rapidly
to give a continuous Schwarzschild limit after the full solution variation is
formed.  Exact angular primitives give
\begin{equation}
 2\pi\int_{-1}^{1}k_\infty^{(m)}\dd x=C_m,\qquad
 2\pi\int_{-1}^{1}k_\infty^{(B)}\dd x=C_B .
\label{eq:density-integrals}
\end{equation}

At any finite radius, the corresponding expressions depend linearly on
\(w=\sqrt{P}\).  Differentiating in \(r\) and using \(w^2=P\) reduces the
closure equation to
\begin{equation}
 \partial_r k_{x\phi}^{(p)}
 -\partial_x k_{r\phi}^{(p)}=0,\qquad p=m,B.
\label{eq:component-closure}
\end{equation}
The radial component vanishes at \(x=\pm1\).  Integration of
\eqref{eq:component-closure} proves equality of every finite-radius integral
with Eq.~\eqref{eq:density-integrals}.  This argument also controls the
nonuniform approach to the limiting density at the poles: the surface
integral is fixed before the pointwise asymptotic expansion is invoked.

\subsection{Closure and field dependence}

For any tangent variation and exact symmetry,
\begin{equation}
 \dd\kform_\xi
 =\omegaf[g;h,\Lie_\xi g]-i_\xi(E^{(1)}[h]\cdot\eps)=0 .
\label{eq:surface-closure}
\end{equation}
Applied to the region bounded by two constant-\(r\) surfaces and the regular
axis segments, Eq.~\eqref{eq:surface-closure} proves radial conservation of
\eqref{eq:charge-one-form}.  On a bifurcation surface,
\(i_\xi\ThetaLW=0\), and the Noether-charge term reduces to
\(\kappa\,\delta A_h/(8\pi)=T\delta S\), yielding
\eqref{eq:horizon-charge}.

For completeness, the horizon differentiation can be performed directly:
\begin{align}
 T\partial_m S&=\frac{1-2B^2m^2}{(1+B^2m^2)^2},&
 T\partial_B S&=-\frac{3Bm^3}{(1+B^2m^2)^2}.
\label{eq:horizon-derivatives}
\end{align}
These are exactly the two coefficients obtained from
\eqref{eq:density-integrals}.  The horizon and limiting boundary therefore
measure the same nonclosed one-form, while their difference at fixed
parameters vanishes by spacetime conservation.

For \(\xi=\mu(m,B)\partial_t\), linearity of the Noether charge gives
\[
 \delta\Q_{\mu\partial_t}
 =\mu\,\delta\Q_{\partial_t}+(\delta\mu)\Q_{\partial_t},
 \qquad
 \Q_{\delta\xi}=(\delta\mu)\Q_{\partial_t}.
\]
The last terms cancel, so the adjusted surface charge is
\begin{equation}
 \kform_{\mu\partial_t}=\mu\kform_{\partial_t}.
\label{eq:field-dependent-scaling}
\end{equation}
This identity underlies both Eqs.~\eqref{eq:barred-charge} and
\eqref{eq:entropy-generator}.

\section{Boundary canonical data and the moving support}
\label{app:boundary}

\subsection{Construction of the Weyl coordinates}

The two-dimensional orbit metric in the \((r,x)\) sector is diagonal,
\begin{equation}
 \gamma_{AB}\dd x^A\dd x^B
 =A\left(\frac{\dd r^2}{f}+\frac{r^2\dd x^2}{X}\right).
\label{eq:orbit-metric}
\end{equation}
Since \(\rho=\Delta_\phi\sqrt{FX}/P\) is harmonic with respect to this
metric, its conjugate \(z\) is determined up to an additive constant.  With
orientation \(\dd r\wedge\dd x>0\), the Cauchy--Riemann equations are
\begin{equation}
 \partial_r z=-\sqrt{\frac{X}{F}}\,\partial_x\rho,
 \qquad
 \partial_x z=\sqrt{\frac{F}{X}}\,\partial_r\rho .
\label{eq:weyl-CR}
\end{equation}
Inserting \(\rho\) and integrating either equation gives the second expression
in Eq.~\eqref{eq:weyl-map}.  Substitution back into both equations verifies
that the integration function is constant.

The limiting embedding follows before any coordinate inversion is attempted.
At fixed \(x\),
\begin{align}
 \frac{P}{r^2}&\longrightarrow B^2A_y,&
 \frac{F}{r^4}&\longrightarrow B^2(1-y^2),\nonumber\\
 X&=(1-x^2)(1+y^2x^2).
\label{eq:weyl-leading-data}
\end{align}
Using \(B=y/[H(1+y^2)^{3/2}]\) and
\(\Delta_\phi=(1+y^2)^{-1}\) then gives
\eqref{eq:weyl-boundary}.  The transformation Jacobian tends to zero on this
surface, so \((\rho,z)\) cease to be local coordinates there.  The parametric
curve remains regular because Eq.~\eqref{eq:weyl-monotonic} never vanishes.
Its topology follows directly: the two endpoints lie on the axis, the
interior has \(\rho>0\), and monotonic \(z(x)\) precludes
self-intersections.

\subsection{Radial symplectic identity}

The variation of the Einstein--Hilbert Lagrangian has the universal form
\begin{equation}
 \delta\bm L=\bm E^{\mu\nu}\delta g_{\mu\nu}
 +\dd\ThetaLW[g;\delta g].
\label{eq:lagrangian-variation}
\end{equation}
On a timelike radial surface with vanishing shift, the pullback of
\(\ThetaLW\) can be reorganized into the canonical potential
\(\Pi^{ab}\delta h_{ab}\), a field-space exact term, and an intrinsic total
derivative.  Antisymmetrizing two variations removes the field-space exact
term.  The intrinsic derivative integrates to zero on a closed
constant-\(t\) section.  Consequently,
\begin{align}
 \left.\omegaf[g;\delta_1g,\delta_2g]\right|_{\Sigma_R}
 ={}&
 \left(\delta_1\Pi^{ab}\delta_2h_{ab}
 -\delta_2\Pi^{ab}\delta_1h_{ab}\right)\nonumber\\
 &\quad{}\times\dd t\wedge\dd x\wedge\dd\phi
 \nonumber\\
 &+\dd_{\Sigma_R}\bm Y(\delta_1g,\delta_2g).
\label{eq:radial-symplectic-identity}
\end{align}
For the regular axes, the corner integral of \(\bm Y\) vanishes.  This proves
the integrated Brown--York/Lee--Wald relation at every finite \(R\).
Substitution of the exact static metric shows that the local \(\bm Y\) term
also vanishes in the material \(x\) representation used in
Eq.~\eqref{eq:BY-LW-match}.

\subsection{Limiting induced data}

On \(\Sigma_R\), write the radial metric in ADM form
\[
 \dd s^2=N_R^2\dd r^2+h_{ab}\dd x^a\dd x^b,\qquad
 N_R=\sqrt{g_{rr}} .
\]
The vanishing shift and diagonal static metric give
\begin{equation}
 K^{\rm out}_{ab}=\frac{1}{2N_R}\partial_r h_{ab},\qquad
 K_{\rm out}=\frac{1}{2N_R}h^{ab}\partial_r h_{ab}.
\label{eq:extrinsic-curvature}
\end{equation}
The finite limiting data become compact after defining
\begin{equation}
 q=\sqrt{1-(1-y^2)x^2},\qquad
 v=q+yx^2,\qquad s=\sqrt{1-y^2}.
\label{eq:boundary-abbreviations}
\end{equation}
In the diagonal ordering \((t,x,\phi)\), the induced metric is
\begin{align}
 h_{tt}^{\infty}
 &=-\frac{s^2v^2}{4q^4},\nonumber\\
 h_{xx}^{\infty}
 &=\frac{H^2(1+y^2)^3v^2}
 {4y^2(1-x^2)(1+y^2x^2)q^4},\nonumber\\
 h_{\phi\phi}^{\infty}
 &=\frac{4H^2(1-x^2)(1+y^2)(1+y^2x^2)}
 {y^2v^2}.
\label{eq:boundary-induced-metric}
\end{align}
The seam-oriented curvature
\(\widehat K^a{}_b=-K^{{\rm out}\,a}{}_b\) has components
\begin{widetext}
\begin{align}
 \widehat K^t{}_t={}&-\frac{2y}
 {Hs(1+y^2)^{3/2}vq}
 \left[
 x^2y^4-x^2y^3q-2x^2y^2+x^2yq+x^2+y^2+yq-1
 \right],\nonumber\\
 \widehat K^x{}_x={}&
 \frac{2ys}
 {H(1+y^2)^{3/2}vq}
 \left[x^2y^2-2x^2yq-x^2+1\right],\nonumber\\
 \widehat K^\phi{}_\phi={}&
 -\frac{2ysq}{H(1+y^2)^{3/2}v}.
\label{eq:boundary-extrinsic-components}
\end{align}
\end{widetext}
These seam-oriented data are finite at both axis endpoints after converting from
the singular \(x\) coordinate to proper distance.  Their zero-mass limit is
nonuniform for the same reason as Eq.~\eqref{eq:massless-annulus}.

For reference, the natural reflected extension gives
\begin{widetext}
\begin{align}
 S^t{}_t={}&-\frac{x^2y^2s}
 {\pi H(1+y^2)^{3/2}v},\nonumber\\
 S^x{}_x={}&
 \frac{y^2(x^2y^2-x^2-1)}
 {2\pi Hs(1+y^2)^{3/2}v},\nonumber\\
 S^\phi{}_\phi={}&-\frac{y}
 {2\pi Hs(1+y^2)^{3/2}vq}
 \left[
 2x^2y^4-3x^2y^3q-4x^2y^2+3x^2yq
 {}+2x^2+2y^2+yq-2
 \right].
\label{eq:full-israel-components}
\end{align}
\end{widetext}
The three entries obey Eq.~\eqref{eq:Israel-stress} identically.  They also
show that the finite-mass source occupies the entire closed surface.

Substitution into Eq.~\eqref{eq:BY-momentum} expresses the canonical data
entirely through the three induced metric components and their radial
derivatives.  Antisymmetrizing two parameter variations eliminates second
variations.  Substitution of Eq.~\eqref{eq:extrinsic-curvature} into
Eq.~\eqref{eq:boundary-response}, followed by the Gauss--Codazzi decomposition
of the Einstein--Hilbert symplectic potential, reduces the result to the
pullback of the Lee--Wald current.  For
\((\delta_1,\delta_2)=(\partial_H,\partial_y)\), the large-\(R\) limit is
Eq.~\eqref{eq:BY-LW-match}; angular integration gives
\eqref{eq:integrated-flux}.

Let \(X^A(x;H,y)=(\rho_\infty,z_\infty)\) be the embedding
\eqref{eq:weyl-boundary}.  A surface distribution with Lagrangian density
\(\mathcal{S}(x;H,y)\) has Eulerian representation
\begin{equation}
\mathcal{S}_{\mathrm E}(\rho,z)
 =\int_{-1}^{1}\mathcal{S}(x)\,
 \delta^{(2)}\!\left[(\rho,z)-X(x)\right]\dd x .
\label{eq:eulerian-source}
\end{equation}
Its parameter variation is
\begin{align}
\delta\mathcal{S}_{\mathrm E}
 =\int_{-1}^{1}\bigl[
 \delta\mathcal{S}\,\delta^{(2)}(\bm{X}_{W}-\bm{X})
 \nonumber\\
 \hspace{14mm}
 -\mathcal{S}\,\delta X^A\partial_A
 \delta^{(2)}(\bm{X}_{W}-\bm{X})
 \bigr]\dd x .
\label{eq:moving-source}
\end{align}
The second term is the displacement of the support.  It is required for
equivalence between the fixed-\(x\) canonical response and its fixed-Weyl-point
description.

\section{Analytic proofs and limiting charts}
\label{app:proofs}

\subsection{Coordinate differentials and tangent basis}

Differentiating the inverse map~\eqref{eq:Hy-inverse} gives
\begin{align}
 \dd m={}&(1+y^2)^{3/2}\dd H
 +3Hy\sqrt{1+y^2}\,\dd y,\nonumber\\
 \dd B={}&-\frac{y}{H^2(1+y^2)^{3/2}}\dd H
 +\frac{1-2y^2}{H(1+y^2)^{5/2}}\dd y .
\label{eq:coordinate-differentials}
\end{align}
Thus the solution-space tangent fields are related by
\begin{align}
 h_H={}&(1+y^2)^{3/2}h_m
 -\frac{y}{H^2(1+y^2)^{3/2}}h_B,\nonumber\\
 h_y={}&3Hy\sqrt{1+y^2}\,h_m
 +\frac{1-2y^2}{H(1+y^2)^{5/2}}h_B .
\label{eq:tangent-basis}
\end{align}
Substitution of Eq.~\eqref{eq:coordinate-differentials} into
\eqref{eq:charge-one-form} cancels the \(\dd y\) coefficient and gives
\eqref{eq:alpha-Hy}.  The same basis change converts the integrated
Lee--Wald current from Eq.~\eqref{eq:charge-curl} to
\eqref{eq:integrated-flux}.

\subsection{Finite homothety and its tangent}

Set \(T=Bt\), \(R=Br\), and \(y=Bm\).  Every occurrence of \(m,r,B\) in the
dimensionless functions \(P,w,u\), and \(f\) then combines into \(y\) and
\(R\), while each coordinate line element contributes the appropriate power
of \(B^{-1}\).  Factoring the common power from Eq.~\eqref{eq:metric} gives
\begin{equation}
 \dd s^2(m,B;t,r,x,\phi)
 =B^{-2}\dd s^2(y,1;T,R,x,\phi),
\label{eq:homothety-line-element}
\end{equation}
which is Eq.~\eqref{eq:homothety}.  Differentiation along a curve with
\(y\) fixed yields
\[
 V=B\partial_B-m\partial_m,\qquad V=-H\partial_H .
\]
The coordinate part of the derivative is generated by
\(D=t\partial_t+r\partial_r\); differentiating the prefactor
\(B^{-2}\) supplies the \(-2g\) term in
\eqref{eq:scaling-variation}.

This same calculation gives the scale weights
\begin{align}
 V(M)&=-M,\qquad V(T)=T,\nonumber\\
 V(S)&=-2S,\qquad V(\Psi_y)=-\Psi_y .
\label{eq:scale-weights}
\end{align}
Contracting the first law~\eqref{eq:global-first-law} with \(V\) reproduces
\(-M=T(-2S)\), equivalent to the Smarr identity.  The nonzero result is the
exact-symmetry charge test used in Sec.~\ref{sec:scaling}.

\subsection{Integrating factors}

Writing a general rescaled charge as
\(\beta=\mu(H,y)\sqrt{1+y^2}\,\dd H\), its exterior derivative is
\begin{equation}
 \dd\beta
 =-\partial_y\!\left[\mu(H,y)\sqrt{1+y^2}\right]
 \dd H\wedge\dd y .
\label{eq:integrating-factor-proof}
\end{equation}
Closure is equivalent to
\(\mu\sqrt{1+y^2}=F(H)\), which proves
\eqref{eq:integrating-factors}.  The boundary value
\(\mu(H,0)=1\) for every \(H>0\) then gives \(F(H)=1\).

\subsection{Monotonicity of the horizon shape}

An integral representation convenient for differentiating
\eqref{eq:shape-ratio} is
\begin{equation}
 R(y)\equiv\frac{C_p}{C_e}
 =\frac{2}{\pi}\int_0^{\pi/2}
 \frac{\sqrt{1+y^2\sin^2\vartheta}}{1+y^2}\,\dd\vartheta .
\label{eq:shape-integral}
\end{equation}
Differentiation under the integral sign yields
\begin{equation}
 R'(y)=-\frac{2y}{\pi(1+y^2)^2}
 \int_0^{\pi/2}
 \frac{2+y^2\sin^2\vartheta-\sin^2\vartheta}
 {\sqrt{1+y^2\sin^2\vartheta}}\,\dd\vartheta<0
\label{eq:shape-monotonic-proof}
\end{equation}
for \(0<y<1\).  Every factor in the integral is positive.  Fixed-\(H\)
variations of \(y\) therefore change an intrinsic dimensionless observable.

\subsection{Degenerate limits}

The Schwarzschild boundary \(y=0\) is regular in \((H,y)\), although the
\(B=1\) homothety chart in Eq.~\eqref{eq:homothety} becomes singular there.
As \(y\to1^-\), the horizon coordinate radius diverges while
\(H\), \(S\), and the normalized charge remain finite when held as phase
coordinates.  At \(y=1/\sqrt2\), the map from \((H,y)\) to \((S,B)\) loses
rank according to Eq.~\eqref{eq:B-turning-map}; Eq.~\eqref{eq:global-first-law}
remains smooth.

For the massless Weyl limit, fixed interior \(x\) gives
\eqref{eq:massless-annulus}, whereas taking \(x\to\pm1\) first keeps each
finite-mass endpoint on the axis.  This noncommutativity converts the compact
finite-mass support into a noncompact annulus and accounts for the inner-edge
boundary layer of Eq.~\eqref{eq:HN-density}.

\subsection{Weyl boundary scales and the annular layer}

Near the north-axis scale in Eq.~\eqref{eq:weyl-scales-main}, set
\(\varepsilon=1-x^2\).  The local expansion is
\begin{align}
 \rho_\infty&=
 \frac{H(1+y^2)\sqrt{1-y^2}}{y^3}\sqrt{\varepsilon}
 +O(\varepsilon^{3/2}),\nonumber\\
 z_\infty&=\frac{H\sqrt{1+y^2}}{y^2}+O(\varepsilon).
\label{eq:axis-boundary-expansion}
\end{align}
This square-root approach is smooth in proper polar coordinates for every
fixed \(0<y<1\).

Along the massless path at fixed \(B\),
\(H=y/[B(1+y^2)^{3/2}]\).  At fixed interior \(x\), expansion of the embedding
gives
\begin{align}
 \rho_\infty={}&\frac{1}{B\sqrt{1-x^2}}
 -\frac{y^2(x^4-2x^2+3)}
 {2B(1-x^2)^{3/2}}+O(y^4),\nonumber\\
 z_\infty={}&\frac{yx}{B(1-x^2)}+O(y^3).
\label{eq:massless-boundary-expansion}
\end{align}
The correction is nonuniform as \(x\to\pm1\), while the leading term is
exactly the annular map~\eqref{eq:massless-annulus}.  Near its inner edge,
\(\rho=B^{-1}+\ell\), the density behaves as
\begin{align}
 \sigma(\rho)
 &=\frac{\sqrt{B}}{2\pi\sqrt{2\ell}}
 +O(\sqrt{\ell}),\nonumber\\
 \sigma(\rho)&=\frac{1}{2\pi\rho}+O(\rho^{-3})
 \quad(\rho\to\infty).
\label{eq:HN-density-asymptotics}
\end{align}
The first singularity is locally integrable.  Together with
\eqref{eq:massless-boundary-expansion}, it identifies the annular source as a
boundary layer of the smooth finite-mass surfaces.

\subsection{Schwarzschild and strong-deformation endpoints}

At fixed \(m\), the Schwarzschild boundary is analytic in \(y^2\).  The
leading expansions of the thermodynamic variables are
\begin{align}
 M&=m\left(1-y^2+y^4+O(y^6)\right),\nonumber\\
 H&=m\left(1-\frac32y^2+\frac{15}{8}y^4+O(y^6)\right),\nonumber\\
 S&=4\pi m^2\left(1-3y^2+6y^4+O(y^6)\right),\nonumber\\
 T&=\frac{1}{8\pi m}\left(1+2y^2+y^4\right).
\label{eq:schwarzschild-expansions}
\end{align}
The horizon shape has the independent expansion
\begin{equation}
 \frac{C_p}{C_e}
 =1-\frac34y^2+\frac{45}{64}y^4+O(y^6).
\label{eq:shape-small-y}
\end{equation}
The absence of a linear term reflects the invariance of the metric under
\(B\mapsto-B\).  The \(B=0\) line still supplies a full normalization
boundary because its mass parameter ranges over all positive values.  This is
why the condition \(\mu(m,0)=1\) determines the function \(F(H)\) in
\eqref{eq:integrating-factors}.

At the other endpoint, \(y\to1^-\), the Schwarzschild-like radial coordinate
of the horizon diverges,
\begin{equation}
 r_h=\frac{2m}{1-y^2}\longrightarrow\infty,
\label{eq:strong-horizon-radius}
\end{equation}
while invariant thermodynamic quantities have finite limits at fixed \(m\):
\begin{align}
 M&\to\frac{m}{2},\qquad
 H\to\frac{m}{2\sqrt2},\nonumber\\
 S&\to\frac{\pi m^2}{2},\qquad
 T\to\frac{1}{2\pi m}.
\label{eq:strong-thermo-limits}
\end{align}
The coordinate divergence therefore marks the edge of the chosen exterior
chart; entropy and normalized energy remain finite.  The Weyl
equatorial radius in Eq.~\eqref{eq:weyl-scales-main} tends to zero at fixed
\(H\), showing that the limiting boundary simultaneously develops a
degenerate shape.  Our open domain \(0<y<1\) keeps both the horizon chart and
the boundary embedding regular.

\subsection{The fixed-entropy source chart}

At fixed \(H\), define \(b(y)=BH=y(1+y^2)^{-3/2}\).  Its derivative is
\begin{equation}
 b'(y)=\frac{1-2y^2}{(1+y^2)^{5/2}}.
\label{eq:b-prime}
\end{equation}
It is positive below \(y_*\), negative above \(y_*\), and vanishes only at
\eqref{eq:B-turning-point}.  The source one-form transforms without a
singularity:
\begin{equation}
 \Psi_y\dd y
 =\frac{\Psi_y}{b'(y)/H}\,\dd B
 =\Phi_B\dd B
\label{eq:source-one-form-transform}
\end{equation}
on either open branch.  At the turning point \(\dd B\) vanishes to first order
along a fixed-\(H\) displacement, so the pole in \(\Phi_B\) is exactly
cancelled in the product.

\bibliography{references}

\end{document}